\documentclass{article}
\usepackage{amssymb}
\usepackage{amsfonts}
\usepackage{amsmath}
\usepackage{epsfig}

\begin{document}

\title{Family symmetries and the SUSY flavour problem}
\author{I. de Medeiros Varzielas\thanks{%
i.varzielas@physics.ox.ac.uk}, G. G. Ross\thanks{%
g.ross@physics.ox.ac.uk} \\
%EndAName
Rudolf Peierls Centre for Theoretical Physics,\\
University of Oxford, 1 Keble Road, Oxford, OX1 3NP}
\maketitle

\begin{abstract}
We re-examine the constraints on continuous family symmetries coming from
flavour changing neutral current limits on the $D-$term contributions to
squark and slepton masses. We show that, for a restricted choice of the
familon sector, continuous family symmetries are consistent with even the
most conservative limits both for the case of gauge mediated supersymmetry
breaking and the case of gravity mediated supersymmetry breaking.
\end{abstract}

\section{Introduction}

The SUSY flavour problem is a long-standing problem for SUSY theories. Due
to the introduction of additional flavoured particles, the sfermions, there
are extra contributions to 4-fermion flavour changing interactions, leading
to flavour changing neutral currents (FCNC) that are potentially too large.
For example, FCNC processes can occur for quarks through box diagrams
involving gluino and squarks in the internal lines, with the contribution
depending on the masses of the squarks that mediate the flavour change.
Considering such processes, experiments have placed rather stringent
constraints on the mass matrices of the sfermions \cite{Masiero}. While
there are alternatives (e.g. alignment of the sfermions with the fermions) 
\cite{Nir}, the simplest way to satisfy the constraints is by requiring the
sfermions of each family to have nearly degenerate masses.

There are three established ways of obtaining the required sfermion mass
degeneracy: gauge mediated models, in which case the SUSY breaking mechanism
giving rise to the soft masses is generation blind; SUGRA models, where
universal soft masses are generated by gravitational interactions; and family
symmetries, where the added symmetry explaining the fermion mass structure
requires the sfermions of each family to have nearly degenerate masses.

Although these mechanisms are necessary to avoid large FCNC they may not be
sufficient if there are further sources of family dependent masses. This is
the case if there is a continuous family symmetry because the associated $D$%
-term \footnote{%
If the family symmetry is discrete there are no $D$-terms associated with it
e.g. \cite{Kubo}. In this case there may be light pseudo Goldstone bosons
associated with the breaking of the family symmetry.} spoils the desired
degeneracy of sfermion masses \cite{Murayama} and is commonly thought to
rule out symmetries which differentiate between the first two families. In
this paper we show how this problem can readily be avoided. The way this
works depends on the origin of the soft masses and we discuss the cases of
both gravity and gauge mediation.

In section \ref{sec:SCKM} we review how to express the sfermion mass
matrices in the \textquotedblleft Super-CKM\textquotedblright\ (SCKM) basis 
\cite{Hall} and compare with the experimental bounds presented in \cite%
{Masiero}. Section \ref{sec:Consider} shows how the $D$-term gives rise to
family dependent contributions which potentially violate the FCNC bounds. We also
present a method for obtaining an upper bound on
the model predictions for the FCNC effects and give a discussion of
the energy scales relevant to the analysis. In section \ref{sec:Conclusion}
we present ways of solving the SUSY flavour problem associated with
continuous family symmetries. We conclude with a summary in Section \ref%
{sec:Summary}.

\section{Super-CKM basis and the experimental bounds \label{sec:SCKM}}

In SUSY models we need to specify not just the basis chosen for the fermion
states, as in the Standard Model (SM), but also the basis chosen for the
sfermion states. In the SM it is often convenient to use either the
\textquotedblleft Mass basis\textquotedblright , where each fermion state
has a well defined mass (the mass matrices are diagonal) and the
\textquotedblleft Weak basis\textquotedblright , where each fermion state
has a well defined flavour (the weak interaction matrix is diagonal).
Similarly, it is convenient to use a specific SUSY basis, and it is
particularly useful to use the SCKM basis \cite{Hall}. We turn to a brief
review of the SCKM basis, illustrating it by the down quark and squark
sector. We generalize the mass matrices to a $6\times 6$ notation,
distinguishing left-handed (LH) and right-handed (RH). In an arbitrary basis
for the quarks, we have:

\begin{equation}
(\bar{d^{\prime }}_{L} , \bar{d^{\prime }}_{R}) \left[ 
\begin{array}{cc}
0 & M_{d}^{D} \\ 
M_{d}^{D^{\dagger}} & 0%
\end{array}%
\right] \left( 
\begin{array}{c}
d^{\prime }_{L} \\ 
d^{\prime }_{R}%
\end{array}
\right)  \label{eq:mdD}
\end{equation}
where $d^{\prime }_{L}$, $d^{\prime }_{R}$ are 3 component columns
containing the 3 down-type quarks ($d$, $s$, $b$), and the $6 \times 6$
matrix is represented in four $3 \times 3$ blocks. The prime denotes that
the states are taken in the arbitrary basis. $M_{d}^{D}$ is the Dirac mass
matrix for the down quarks.

The associated scalar partners have a squared mass matrix containing their
squared masses:

\begin{equation}
M_{\tilde{d}}^{2}=\left[ 
\begin{array}{cc}
M_{\tilde{d}}^{LL^{2}} & M_{\tilde{d}}^{LR^{2}} \\ 
M_{\tilde{d}}^{RL^{2}} & M_{\tilde{d}}^{RR^{2}}%
\end{array}%
\right]  \label{eq:smass}
\end{equation}%
where we expressed the full $6\times 6$ matrix in terms of $3\times 3$ block
matrices. In the LR quadrant, $M_{\tilde{d}}^{LR^{2}}$ is equal to the down
quark Dirac mass matrix $M_{d}^{D}$ times a generation independent mass
factor, and in the RL quadrant, $M_{\tilde{d}}^{RL^{2}}$ is similarly
proportional to the hermitian conjugate, $M_{d}^{D^{\dagger }}$.

In the LL and RR quadrants we have squark masses for the LH and RH squarks
respectively. We start in a basis where these are diagonal, and parametrize
the matrix with an explicit universal contribution $m_{0}^{2}$ that is
generation blind, plus deviations from the degenerate spectrum which are
parametrized as $\Delta m_{\tilde{f}}^{2}$. It is useful to use such a
parametrization as we will be considering models where the common $m_{0}$
comes from a specific SUSY breaking messenger mechanism (e.g. SUGRA), and
deviations arise from $D$-term contributions associated with a continuous
family symmetry. For the down-type squarks we have:

\begin{equation}
M_{\tilde{d}}^{LL;RR^{2}} = \left[ 
\begin{array}{ccc}
m_{0}^{2} + \Delta m_{\tilde{d}_{L;R}}^{2} & 0 & 0 \\ 
0 & m_{0}^{2} + \Delta m_{\tilde{s}_{L;R}}^{2} & 0 \\ 
0 & 0 & m_{0}^{2} + \Delta m_{\tilde{b}_{L;R}}^{2}%
\end{array}
\right]  \label{eq:squark mass}
\end{equation}

It is convenient to re-express this matrix in the SCKM basis. The SCKM basis
consists of having the fermion states in the basis where their Dirac mass
matrices are diagonalised, and the sfermion states in the basis that has the
neutral gauginos couplings flavour diagonal. This requires changing the
quark and squark states by the same transformations. To do this we use the $%
6\times 6$ mixing matrix that diagonalises the quark Dirac masses:

\begin{equation}
\left( 
\begin{array}{c}
d_{L} \\ 
d_{R}%
\end{array}
\right) = \left[ 
\begin{array}{cc}
V_{L} & 0 \\ 
0 & V_{R}%
\end{array}%
\right] \left( 
\begin{array}{c}
d^{\prime }_{L} \\ 
d^{\prime }_{R}%
\end{array}
\right)
\end{equation}
where the unprimed quark states are the mass eigenstates. From eq.(\ref%
{eq:mdD}), we get:

\begin{equation}
(\bar{d}_{L},\bar{d}_{R})\left[ 
\begin{array}{cc}
V_{L} & 0 \\ 
0 & V_{R}%
\end{array}%
\right] \left[ 
\begin{array}{cc}
0 & M_{d}^{D} \\ 
M_{d}^{D^{\dagger }} & 0%
\end{array}%
\right] \left[ 
\begin{array}{cc}
V_{L}^{\dagger } & 0 \\ 
0 & V_{R}^{\dagger }%
\end{array}%
\right] \left( 
\begin{array}{c}
d_{L} \\ 
d_{R}%
\end{array}%
\right)
\end{equation}%
where the product of the three $6\times 6$ matrices will result in
diagonalised LR and RL quadrants (the diagonal down quark Dirac mass matrix
and its hermitian conjugate, respectively). In the SCKM basis the down
squark mass matrix has the form:

\begin{equation}
M_{\tilde{d}}^{2} = \left[ 
\begin{array}{cc}
V_{L} M_{\tilde{d}}^{LL^{2}} V_{L}^{\dagger} & V_{L} M_{\tilde{d}}^{LR^{2}}
V_{R}^{\dagger} \\ 
V_{R} M_{\tilde{d}}^{RL^{2}} V_{L}^{\dagger} & V_{R} M_{\tilde{d}}^{RR^{2}}
V_{R}^{\dagger}%
\end{array}%
\right]  \label{eq:sdownLL}
\end{equation}

Note that the LR and RL blocks are now diagonal, but (due to the $\Delta m_{%
\tilde{f}}^{2}$) the LL and RR blocks need not be.

The mass insertions $\Delta$ \cite{Masiero} are defined as the components of
the sfermion mass matrix in the SCKM basis. For example, $\Delta^{\tilde{d}%
}_{12_{LL}}$ (the $\tilde{d}_{L} \tilde{s}_{L}$ component) is given by:

\begin{equation}
\Delta^{\tilde{d}}_{12_{LL}} \equiv (m_{0}^2 + \Delta m_{\tilde{d}_{L}}^{2})
V_{L_{11}} V_{L_{21}}^{*} + (m_{0}^2 + \Delta m_{\tilde{s}_{L}}^{2})
V_{L_{12}} V_{L_{22}}^{*} + (m_{0}^2 + \Delta m_{\tilde{b}_{L}}^{2})
V_{L_{13}} V_{L_{23}}^{*}  \label{eq:m_12}
\end{equation}

Since $V_{L}$ is unitary:

\begin{equation}
V_{L_{11}} V_{L_{21}}^{*} + V_{L_{12}} V_{L_{22}}^{*} + V_{L_{13}}
V_{L_{23}}^{*} = 0  \label{eq:unitary}
\end{equation}

Using this immediately shows that the terms proportional to $m_{0}^{2}$
contribution in eq.(\ref{eq:m_12}) vanish as expected (as would any
generation independent contribution).

In \cite{Masiero} the experimental constraints are presented in terms of
quantities $\delta $, which are obtained by dividing the mass insertions $%
\Delta $ by the average sfermion mass. To illustrate, with down squarks in
the LL block, we have (from eq.(\ref{eq:sdownLL}), eq.(\ref{eq:m_12}), eq.(%
\ref{eq:unitary})):

\begin{eqnarray}
\delta _{12_{LL}}^{d} &\equiv &\frac{\left( V_{L}M_{\tilde{d}%
}^{LL^{2}}V_{L}^{\dagger }\right) _{12}}{\langle m_{\tilde{q}}^{2}\rangle } 
\notag \\
&=&\frac{\Delta m_{\tilde{d}_{L}}^{2}V_{L_{11}}V_{L_{21}}^{\ast }+\Delta m_{%
\tilde{s}_{L}}^{2}V_{L_{12}}V_{L_{22}}^{\ast }+\Delta m_{\tilde{b}%
_{L}}^{2}V_{L_{13}}V_{L_{23}}^{\ast }}{\langle m_{\tilde{q}}^{2}\rangle }
\label{eq:delta_12}
\end{eqnarray}%
where $\langle m_{\tilde{q}}^{2}\rangle $ is the geometrical average for the
squark mass \cite{Masiero}.

The most stringent experimental upper bounds for the $\delta $ from \cite%
{Masiero} are shown in Table \ref{ta:bounds} (for quarks) and Table \ref%
{ta:lbounds} (for leptons) \footnote{%
Making some assumptions about the underlying physics (e.g. GUTs) one may
obtain stronger bounds \cite{Masiero} but we do not consider these here.}.

% Here are the tables with the results from the reference

\begin{table}[htbp]
\centering
\begin{tabular}{|c|c|c|}
\hline
$\frac{m_{\tilde{g}}^{2}}{m_{\tilde{q}}^{2}}$ & $\sqrt{\left| \mathrm{Re}
\left( \delta^{d}_{12_{LL}} \right)^{2} \right|}$ & $\sqrt{\left| \mathrm{Re}
\delta^{d}_{12_{LL}} \delta^{d}_{12_{RR}} \right|}$ \\ \hline
$0.3$ & $1.9 \times 10^{-2}$ & $2.5 \times 10^{-3}$ \\ 
$1.0$ & $4.0 \times 10^{-2}$ & $2.8 \times 10^{-3}$ \\ 
$4.0$ & $9.3 \times 10^{-2}$ & $4.0 \times 10^{-3}$ \\ \hline
$\frac{m_{\tilde{g}}^{2}}{m_{\tilde{q}}^{2}}$ & $\sqrt{\left| \mathrm{Re}
\left( \delta^{d}_{13_{LL}} \right)^{2} \right|}$ & $\sqrt{\left| \mathrm{Re}
\delta^{d}_{13_{LL}} \delta^{d}_{13_{RR}} \right|}$ \\ \hline
$0.3$ & $4.6 \times 10^{-2}$ & $1.6 \times 10^{-2}$ \\ 
$1.0$ & $9.8 \times 10^{-2}$ & $1.8 \times 10^{-2}$ \\ 
$4.0$ & $2.3 \times 10^{-1}$ & $2.5 \times 10^{-2}$ \\ \hline
$\frac{m_{\tilde{g}}^{2}}{m_{\tilde{q}}^{2}}$ & $\sqrt{\left| \mathrm{Re}
\left( \delta^{u}_{12_{LL}} \right)^{2} \right|}$ & $\sqrt{\left| \mathrm{Re}
\delta^{u}_{12_{LL}} \delta^{u}_{12_{RR}} \right|}$ \\ \hline
$0.3$ & $4.7 \times 10^{-2}$ & $1.6 \times 10^{-2}$ \\ 
$1.0$ & $1.0 \times 10^{-1}$ & $1.7 \times 10^{-2}$ \\ 
$4.0$ & $2.4 \times 10^{-1}$ & $2.5 \times 10^{-2}$ \\ \hline
$\frac{m_{\tilde{g}}^{2}}{m_{\tilde{q}}^{2}}$ & $\sqrt{\left| \mathrm{Im}
\left( \delta^{d}_{12_{LL}} \right)^{2} \right|}$ & $\sqrt{\left| \mathrm{Im}
\delta^{d}_{12_{LL}} \delta^{d}_{12_{RR}} \right|}$ \\ \hline
$0.3$ & $1.5 \times 10^{-3}$ & $2.0 \times 10^{-4}$ \\ 
$1.0$ & $3.2 \times 10^{-3}$ & $2.2 \times 10^{-4}$ \\ 
$4.0$ & $7.5 \times 10^{-3}$ & $3.2 \times 10^{-4}$ \\ \hline
$\frac{m_{\tilde{g}}^{2}}{m_{\tilde{q}}^{2}}$ & \multicolumn{2}{|c|}{$\left|
\delta^{d}_{23_{LL}} \right|$} \\ \hline
$0.3$ & \multicolumn{2}{|c|}{$4.4$} \\ 
$1.0$ & \multicolumn{2}{|c|}{$8.2$} \\ 
$4.0$ & \multicolumn{2}{|c|}{$26$} \\ \hline
\end{tabular}%
\caption{Bounds for $\protect\delta$, assuming $m_{\tilde{q}} = 500$ GeV 
\protect\cite{Masiero}}
\label{ta:bounds}
\end{table}

\begin{table}[htbp]
\centering
\begin{tabular}{|c|c|}
\hline
$\frac{m_{\tilde{\gamma}}^{2}}{m_{\tilde{l}}^{2}}$ & $\left|
\delta^{e}_{12_{LL}} \right|$ \\ \hline
$0.3$ & $4.1 \times 10^{-3}$ \\ 
$1.0$ & $7.7 \times 10^{-3}$ \\ 
$5.0$ & $3.2 \times 10^{-2}$ \\ \hline
$\frac{m_{\tilde{\gamma}}^{2}}{m_{\tilde{l}}^{2}}$ & $\left|
\delta^{e}_{13_{LL}} \right|$ \\ \hline
$0.3$ & $15$ \\ 
$1.0$ & $29$ \\ 
$5.0$ & $1.2 \times 10^{2}$ \\ \hline
$\frac{m_{\tilde{\gamma}}^{2}}{m_{\tilde{l}}^{2}}$ & $\left|
\delta^{e}_{23_{LL}} \right|$ \\ \hline
$0.3$ & $2.8$ \\ 
$1.0$ & $5.3$ \\ 
$5.0$ & $22 $ \\ \hline
\end{tabular}%
\caption{Bounds for $\protect\delta$, assuming $m_{\tilde{l}} = 100$ GeV 
\protect\cite{Masiero}}
\label{ta:lbounds}
\end{table}

\section{The family symmetry flavour problem \label{sec:Consider}}

\subsection{$D$-term contributions \label{sec:Dterms}}

We now illustrate how $D$-terms associated with continuous family symmetry
groups can generate family dependant contributions to the sfermion masses.
We consider a simple example of $U(1)_{family}$ symmetry, with the field
content extended to include two familons, $\phi $ and $\bar{\phi}$. We
define the coupling constant so that the charge of $\phi $ is +1, meaning
that the other family charges are defined relative to the charge of this
familon. The $D$-term associated with the continuous family symmetry is then:

\begin{equation}
\left( D-\mathrm{term} \right) ^{2} = g_{f}^{2} \left( |\phi|^{2} + c |\bar{%
\phi}|^{2} + c_{\tilde{d}_{L}} \tilde{d}_{L}^{2} + c_{\tilde{d}_{R}} \tilde{d%
}_{R}^{2} + (...) \right)^{2}
\end{equation}
where $g_{f}$ is the family coupling constant, $c$ is the family charge of $%
\bar{\phi}$, $c_{\tilde{d}_{L;R}}$ are the family charges of the down squark
with left and right handedness respectively, and the (...) stands for
similar terms for all the other sfermions.

Expanding the $D$-term we can identify terms quadratic in the down squarks,
i.e. potential contributions for their masses. Using the notation of eq.(\ref%
{eq:squark mass}):

\begin{equation}
\Delta m_{\tilde{f}_{L;R}}^{2}=2 c_{\tilde{f}_{L;R}}\left\langle
D^{2}\right\rangle  \label{eq:Delta_m}
\end{equation}%
where%
\begin{equation*}
\left\langle D^{2} \right\rangle = g_{f}^{2} \langle |\phi |^{2}+c|\bar{\phi}%
|^{2}\rangle
\end{equation*}%
is the magnitude of the $D-$term. The contributions shown in eq.(\ref%
{eq:Delta_m}) are generation dependant, and give rise to the family symmetry
flavour problem. We now quantify the problem by calculating the $\delta$
predicted by the model, allowing for a direct comparison with the
experimental bounds. For example, we substitute eq.(\ref{eq:Delta_m}) into
eq.(\ref{eq:delta_12}):

\begin{equation}
\delta _{12_{LL}}^{d}\simeq \frac{2 \left\langle D^{2}\right\rangle (c_{%
\tilde{d}_{L}}V_{L_{11}}V_{L_{21}}^{\ast }+c_{\tilde{s}%
_{L}}V_{L_{12}}V_{L_{22}}^{\ast }+c_{\tilde{b}_{L}}V_{L_{13}}V_{L_{23}}^{%
\ast })}{\langle m_{\tilde{q}}^{2}\rangle }  \label{eq:cdelta_12}
\end{equation}

Similar expressions are obtained for the other $\delta$.

\subsection{Upper bounds on the theoretical predictions \label{sub:Unknown}}

Since the mixing matrices are a priori unknown, it is useful to derive an
upper bound on the $\delta$ that is independent of them. Consider just the
part of eq.(\ref{eq:cdelta_12}) dependant on the charges and on the mixing
matrix entries:

\begin{equation}
c_{\tilde{d}_{L}} V_{L_{11}} V_{L_{21}}^{*} + c_{\tilde{s}_{L}} V_{L_{12}}
V_{L_{22}}^{*} + c_{\tilde{b}_{L}} V_{L_{13}} V_{L_{23}}^{*}
\label{eq:mixing}
\end{equation}

Each of the terms in eq.(\ref{eq:mixing}) contains two elements of the
mixing matrix that share a common column (e.g. $V_{L_{11}}V_{L_{21}}^{\ast }$%
). We designate these as \textquotedblleft mixing pairs\textquotedblright .
The unitarity of the mixing matrix imposes restrictions on the mixing pairs:
for example eq.(\ref{eq:unitary}) has three such pairs summing up to zero (a
unitarity triangle). Also because of unitarity, a mixing pair can be written in the form $\frac{1}{2} \sin (2 \theta) \cos (\phi)$,
where $\theta$ and $\phi$ are mixing angles, thus the maximum magnitude of any
of these pairs is $\frac{1}{2}$.
Furthermore eq.(\ref{eq:unitary}) shows
that if one of the pairs has the maximum magnitude of $\frac{1}{2}$, the
other two pairs have to point opposite in relation to the maximum magnitude
pair (thus closing the respective unitarity triangle). Because of this, in
order to maximize eq.(\ref{eq:mixing}), we identify which two of the three
family charges produce $\mathrm{Max}\left\vert c_{i}-c_{j}\right\vert $ (the
dominant charge difference). To obtain the maximum value the mixing pair
corresponding to the other charge must vanish, and the mixing pairs corresponding
to the other two charges must take the maximum magnitude of $\frac{1}{2}$.

Thus, for example, from eq.(\ref{eq:cdelta_12}):

\begin{equation}
\left\vert \delta _{12}^{LL}\right\vert <\left\vert \frac{\left\langle
D^{2}\right\rangle }{\langle m_{\tilde{q}}^{2}\rangle }\right\vert \mathrm{%
Max}\left\vert c_{i}-c_{j}\right\vert  \label{eq:maximal}
\end{equation}

A specific model can saturate the upper bounds given in eq.(\ref{eq:maximal}%
) if there is maximal mixing in two families with no mixing from the other
family and the families that mix correspond to those that maximise $%
\left\vert c_{i}-c_{j}\right\vert $. Further, when comparing with the
experimental bounds, the most stringent constraints arise if it is the $%
(1,2) $ families that mix corresponding to the most stringent experimental
upper bounds. Finally the phase should also be in the direction that yields
the strictest experimental bound. Given this it is quite unlikely that a
specific model will indeed saturate the bound so the bounds should be
considered as very conservative.

\subsection{Running effects \label{sub:Running}}

Before comparing theory to experiment it is important to discuss the energy
scales at which the comparison should be made. The soft SUSY\ breaking
masses are generated at a scale corresponding to the mediator scale $M_{X}$
communicating SUSY breaking from the hidden to the visible sector and
radiative corrections to the mass will be cutoff at this scale. For the case
of SUGRA this is the Planck scale and there are substantial radiative
corrections in continuing to the electroweak scale where the experimental
bounds are obtained. For the case of gauge mediation the gauge messenger
scale can be much lower than the Planck scale and so the radiative
corrections may be much smaller. The dominant radiative corrections are due
to the gauge interactions which are flavour blind. They have the effect of
increasing $\langle m_{\tilde{q}}^{2}\rangle $ while leaving $\Delta m_{%
\tilde{f}_{L;R}}^{2}$ unchanged. As a result they systematically reduce the
FCNC effects \cite{Ellis}.

Actually it is more convenient, when comparing with the theoretical
expectation, to make the comparison at the messenger scale by continuing the
experimental bounds up in energy. Due to the radiative corrections just
discussed $\delta $ will depend on the scale $\mu $ at which the comparison
is to be made, $\delta =\delta (\mu ^{2})$. We have: 
\begin{equation}
\delta (M_{X}^{2})=\delta (M_{W}^{2})\frac{m_{\tilde{f}}^{2}(M_{W}^{2})}{m_{%
\tilde{f}}^{2}(M_{X}^{2})}  \label{eq:deltap}
\end{equation}

To evaluate the size of the effect, one can use the renormalisation group
equations \cite{Running}. In Table \ref{ta:compare} we display sample values of $\delta(M_{X})$ for gauge and gravity mediation.
For gravity mediation, we considered the low energy squark masses of $m_{\tilde{q}}=500$ GeV and
slepton masses of $m_{\tilde{l}}=100$ GeV (as used in the bounds of \cite%
{Masiero}) and running effects corresponding to a common unified gaugino mass $m_{1/2} \sim
250$ GeV lead the sfermions masses to run to a unified value $%
m_{0}=80$ GeV at the Planck scale ($M_{X}=M_{P}$).
For gauge mediation, we considered the messenger scale to be $M_{X}=200$ TeV (we use the SPS $8$ scenario in \cite{Running}).
The slepton masses don't run significantly up to that energy range; the low scale average squark mass however is considerably higher than in the gravity mediation scenario,
taking the value of $1100$ GeV and running to $1000$ GeV at the messenger scale $M_{X}=200$ TeV. $\delta(M_{W})$ needs to be scaled with respect to the higher squark mass \cite{Masiero} before applying  eq.(\ref{eq:deltap}).

In obtaining the values, we use as starting point the $\delta (M_{W}^{2})$ corresponding to
the mass ratios $\frac{m_{\tilde{\gamma}}^{2}}{m_{\tilde{q}}^{2}}$ and $%
\frac{m_{\tilde{\gamma}}^{2}}{m_{\tilde{l}}^{2}}$ of 1.0 in Tables \ref%
{ta:bounds} and \ref{ta:lbounds} (i.e. the middle rows in each sector).
The bounds for $|\delta |$ shown in Table \ref{ta:compare} are obtained
under the most conservative assumption about the phases to make the bound as
strong as possible. When
two different $\delta $ are present in the original experimental bound (as
in the 2nd column of Table \ref{ta:bounds}), we took the value for $|\delta
_{LL}|$ to be the same as $|\delta _{RR}|$ (this leads to the same upper
bound for $\delta _{LL}$ and $\delta _{RR}$ in consecutive rows of Table \ref%
{ta:compare}).

\begin{table}[t]
\centering
\begin{tabular}{|c|c|c|}
\hline
$\delta(M_{X}^{2})$ & Gauge mediation ($M_{X}=200$ TeV) & Gravity mediation ($M_{X}=M_{P}$) \\ 
\hline
&  &  \\ 
$\left| \delta ^{\prime d}_{12_{LL}} \right|$ & $5.9 \times 10^{-4}$ & $8.6
\times 10^{-3}$ \\ 
&  &  \\ 
$\left| \delta ^{\prime d}_{12_{RR}} \right|$ & $5.9 \times 10^{-4}$ & $8.6
\times 10^{-3}$ \\ 
&  &  \\ 
$\left| \delta ^{\prime d}_{13_{LL}} \right|$ & $4.8 \times 10^{-2}$ & $7.0
\times 10^{-1}$ \\ 
&  &  \\ 
$\left| \delta ^{\prime d}_{13_{RR}} \right|$ & $4.8 \times 10^{-2}$ & $7.0
\times 10^{-1}$ \\ 
&  &  \\ 
$\left| \delta ^{\prime u}_{12_{LL}} \right|$ & $4.5 \times 10^{-2}$ & $6.6
\times 10^{-1}$ \\ 
&  &  \\ 
$\left| \delta ^{\prime u}_{12_{RR}} \right|$ & $4.5 \times 10^{-2}$ & $6.6
\times 10^{-1}$ \\ 
&  &  \\ 
$\left| \delta ^{\prime d}_{23_{LL}} \right|$ & $22$ & $3.2 \times 10^{2}$
\\ 
&  &  \\ 
$\left| \delta ^{\prime e}_{12_{LL}} \right|$ & $7.7 \times 10^{-3}$ & $1.2
\times 10^{-2}$ \\ 
&  &  \\ 
$\left| \delta ^{\prime e}_{13_{LL}} \right|$ & $29$ & $45$ \\ 
&  &  \\ 
$\left| \delta ^{\prime e}_{23_{LL}} \right|$ & $5.3$ & $8.3$ \\ 
&  &  \\ \hline
\end{tabular}%
\caption{Experimental upper bounds evaluated at the mediator scale relevant
to gauge and gravity mediation.}
\label{ta:compare}
\end{table}

It may be seen from Table \ref{ta:compare} that the most stringent limits
apply to the first two generations. For them it is necessary that there
should be a strong suppression of the flavour changing SUSY mass squared
difference compared to the the average squark or slepton mass squared. This
is the family symmetry flavour problem. One should note however that we are
considering the most pessimistic case (as discussed subsection \ref%
{sub:Unknown}). For example, it is quite conceivable that the mixing
matrices feature small mixing, which implies the mixing pairs in e.g. eq.(%
\ref{eq:cdelta_12}) can readily take values $O(10^{-1})$ or even smaller,
rather than $\frac{1}{2}$. How much suppression one allows from the mixing
depends on what is considered natural - requiring them to be very small in
order to completely solve the problem requires alignment \cite{Nir},
which is only natural if explained by some specific mechanism. In this paper
we eschew this explanation and look for a more general explanation for the
suppression of the FCNC.

\section{Solving the family symmetry flavour problem \label{sec:Conclusion}}

In this section we discuss the conditions for the $D-$term in eq.(\ref%
{eq:maximal}) to be anomalously small. The $D-$term is fixed when minimising
the familon potential and this relates it to the familon masses. We illustrate
the general expectation for this in the context of a $U(1)$ family symmetry
with a simple familon sector. As we shall discuss, the important aspects of
the form of the $D-$term are common to more complicated familon sectors and
to non-Abelian family symmetries.

\subsection{$F$-term breaking \label{sub:Fterm}}

We first consider a case where the familons acquire non-vanishing VEVs due
to an $F$-term. We take the real potential to be:

\begin{equation}
V=g_{f}^{2}\left( |\phi |^{2}+c|\bar{\phi}|^{2}\right) ^{2}+m^{2}|\phi |^{2}+%
\bar{m}^{2}|\bar{\phi}|^{2}+g\left\vert \phi \bar{\phi}-M^{2}\right\vert ^{2}
\end{equation}%
where we have the $D$-term, familon soft mass terms and also an $F$-term
coming from a super-potential term $g\left( \phi \bar{\phi}-M^{2}\right)
\psi $. By minimising the potential we find:

\begin{eqnarray}
\left\langle D^{2}\right\rangle &\equiv &\langle g_{f}^{2}\left( |\phi
|^{2}+c|\bar{\phi}|^{2}\right) \rangle \simeq \langle \frac{-m^{2}-\bar{m}%
^{2}/c}{4}\rangle  \label{eq:Dvev} \\
\langle |\phi |^{2} \rangle &\simeq &-c \langle |\bar{\phi}|^{2} \rangle
\end{eqnarray}%
where $c<0$ and we have assumed the family symmetry breaking scale is much
greater than the SUSY soft masses.

\subsection{Radiative breaking \label{sub:Radiative}}

As a second example, we consider a case where the VEVs are driven
radiatively. We take $V$ to have the form:

\begin{equation}
V=g_{f}^{2}\left( |\phi |^{2}+c|\bar{\phi}|^{2}\right) ^{2}+\alpha _{\phi
}|\phi |^{2}m^{\prime 2}\mathrm{ln}\left( \frac{|\phi |^{2}}{|\Lambda |^{2}}%
\right) +\alpha _{\bar{\phi}}|\bar{\phi}|^{2}\bar{m}^{\prime 2}\mathrm{%
ln}\left( \frac{|\bar{\phi}|^{2}}{\bar{|\Lambda |}^{2}}\right)
\end{equation}%
where the two last terms include the effects of radiative corrections, $%
\alpha _{\phi }$ and $\alpha _{\bar{\phi}}$ are the fine structure constants
associated with the interactions of $\phi $ and $\bar{\phi }$ and the
tree level contributios have been absorbed in $\Lambda $ and $\bar{%
\Lambda }$. This gives:

\begin{eqnarray}
\left\langle D^{2}\right\rangle &\equiv &\langle g_{f}^{2}\left( |\phi
|^{2}+c|\bar{\phi}|^{2}\right) \rangle  \notag \\
&\simeq &\langle \frac{-\alpha _{\phi }m^{\prime 2}\mathrm{ln}\left( \frac{%
|\phi |^{2}}{|\Lambda |^{2}}\right) -\alpha _{\bar{\phi}}\bar{m}%
^{\prime 2}\mathrm{ln}\left( \frac{|\bar{\phi}|^{2}}{\bar{|\Lambda |}^{2}}%
\right) /c}{4}\rangle  \label{drunning}
\end{eqnarray}%
which has the form of eq.(\ref{eq:Dvev}), where $m$ and $\bar{m}$ are now
interpreted as running masses:

\begin{equation*}
m^{2}\equiv \alpha _{\phi } m^{\prime 2}\mathrm{ln}\left( \frac{|\phi |^{2}}{|\Lambda |^{2}}%
\right)
\end{equation*}

\begin{equation}
\bar{m}^{2}\equiv \alpha _{\bar{\phi}} \bar{m}^{\prime 2}\mathrm{ln}\left( \frac{|\bar{\phi}%
|^{2}}{\bar{|\Lambda |}^{2}}\right)  \label{eq:running_barm}
\end{equation}

With this form of the $D-$term we have from eq.(\ref{eq:Delta_m}):

\begin{equation}
\Delta m_{\tilde{f}_{L;R}}^{2}\simeq \frac{c_{\tilde{f}_{L;R}}}{2}(-m^{2}-%
\bar{m}^{2}/c)  \label{eq:D_combine}
\end{equation}
which immediately allows us to determine $\delta $ from eq.(\ref{eq:maximal}%
). For example the $(1,2)$ element is

\begin{equation}
\left\vert \delta _{12}^{LL}\right\vert <\left\vert \frac{m^{2}+\bar{m}%
^{2}/c}{\langle m_{\tilde{q}}^{2}\rangle }\right\vert \frac{\mathrm{Max}%
\left\vert c_{i}-c_{j}\right\vert }{4}  \label{eq:fmax}
\end{equation}%
where $\langle m_{\tilde{q}}^{2}\rangle $ is to be evaluated at the mediator
scale. To estimate how large the factor involving the familon masses is we
must consider the origin of the soft masses $m$ and $\bar{m}.$ To do this we
consider separately the case of gravity and of gauge mediation.

\subsection{Gravity mediated SUSY breaking \label{sub:SUGRA}}

We first consider SUGRA as the origin of the familon and sfermion masses.
The soft masses are generated at the Planck scale, so we use the estimated
values in the third column of Table \ref{ta:compare}. It may be seen that the
bounds are only significant for the mixing between the first two
generations. Are there ways these bounds can naturally be satisfied without
appealing to small mixing angles?

As we have stressed, SUGRA models solve the FCNC problem by taking all the
soft masses to have a common value at the Planck scale, arguing that gravity
is family and flavour blind. This naturally extends to the familon sector
too so we expect $m^{2}=\bar{m}^{2}$ at the Planck scale. It is immediately
obvious that, provided there are not large radiative corrections, this
offers an elegant solution to the family symmetry flavour problem too if the
familons have equal but opposite charges, i.e. $c=-1$. In this case, c.f. eq(%
\ref{eq:Dvev}), the $D-$term vanishes and the FCNC bounds are satisfied. The
underlying reason for this is because the familon potential is symmetric
under the interchange of $\phi $ and $\bar{\phi}$. Of course radiative
corrections involving Yukawa couplings may spoil this symmetry but these
radiative corrections are suppressed by the loop correction factor and so
can satisfy the bounds even making the very conservative assumptions about
mixing angles and phases discussed above. Although we illustrated the
generation of a $D-$term by a very simple familon sector the solution
applies too in the case where several familons contribute significantly to the $D-$%
term, provided all their charges have equal magnitude. In this case one may
readily check that the $D-$term still vanishes.

The case of radiative family symmetry breaking is perhaps more interesting
as it does not require the introduction of the mass scale $\mu $. For the
case $c=-1,$ $m^{\prime 2}=\bar{m}^{\prime 2}$ the initial tree level
contributions cancel and we have: 
\begin{equation*}
\left\langle D^{2}\right\rangle =(\alpha _{\phi }-\alpha _{\bar{\phi}%
})m^{\prime 2}\ln (\frac{\left\langle |\phi |^{2}\right\rangle }{M_{P}^{2}})
\end{equation*}%
Since the radiative breaking mechanism requires that the radiative
corrections to the soft masses are of the same order as the tree level
contributions each of the two terms is of $O(m^{\prime 2}).$ Thus, if the $%
D-$term is to vanish, it is necessary for $\alpha _{\phi }=\alpha _{\bar{\phi%
}}$ corresponding to the couplings driving the radiative breaking being
symmetric under the interchange of $\phi $ and $\bar{\phi}$ . The
possibility of such a symmetry is not unnatural for the class of family
models discussed in \cite{ivo} in which the fermion mass hierarchy is
generated through the Froggatt-Nielsen mechanism through the coupling of $%
\phi $ to heavy supermultiplets which come in vectorlike pairs, $X,\bar{%
X}$ and $Y,\bar{Y}$ say. Being vectorlike if the coupling $\phi X%
\bar{Y}$ is allowed then so too is the coupling $\bar{\phi }%
\bar{X}Y$ so it is easy to implement a symmetry connecting these terms.

The case of anomaly mediated SUSY breaking offers another way of
suppressing the FCNC effects because the soft masses are given in terms of
the anomalous dimensions of the fields. If $c=-1$ the gauge contributions to 
$m$ and $\bar{m}$ are equal. The family dependent non-gauge contributions
are expected to be small leading to a suppression of the $D-$term as
discussed above. The important point is that UV effects decouple in anomaly
mediation meaning that the anomalous dimension has only contributions from
fields light at the relevant scale, which here is the scale of family
symmetry breaking. Provided the family dependent couplings of the familons
involve only states heavier than this scale they will not split the
degeneracy driven by the gauge coupling. This is the case in the class of
family models discussed in \cite{ivo} because there the vectorlike
supermultiplet mass, $M,$ is necessarily heavier than the familon VEV to
generate the small expansion parameter $\left\langle \phi \right\rangle /M$
which drives the fermion mass hierarchy.

Yet another way of suppressing the $D-$term is provided by orbifold
compactification of string models where the soft masses depend on the the
modular weights of the superfields \cite{casas} and can be anomalously small
if their modular weights are -3. Thus if the familons have this modular
weight and the squarks and sleptons do not the factor $(m^{2}+\bar{m}%
^{2}/c)/\langle m_{\tilde{q}}^{2}\rangle $ appearing in eq.(\ref{eq:fmax})
may be very small leading to the required $D-$term suppression.

Our discussion so far has dealt with the form of the $D-$term coming from an
Abelian family symmetry. However it also applies to non-Abelian family
symmetries. This may readily be seen from the fact that it is always
possible to choose a basis in which the dominant $D-$term contribution to
the mixing between two particular generations corresponds to a diagonal
generator and thus has the same form as the Abelian case.

\subsection{Gauge mediated SUSY breaking \label{sub:Gauge}}

We turn now to the case where the soft masses are due to gauge mediated SUSY
breaking. Since the mediator mass is low the radiative corrections discussed
in subsection \ref{sub:Running} are small. This may be seen by the values in
the second column of Table \ref{ta:compare} where the bounds are
close to the experimental values obtained at the electroweak scale. To be
consistent with these bounds requires a larger suppression to come from the $%
(m^{2}+\bar{m}^{2}/c)/\langle m_{\tilde{q}}^{2}\rangle $ factor than in the
gravity mediated case.

Fortunately, gauge mediated models naturally provide such a suppression
provided the familons have no direct coupling to the SUSY breaking sector.
This follows because the gauginos do not couple directly to the familons
(the familons are not charged under the SM gauge group) and so the
contributions to the familon masses occur at one loop order higher than the
contributions to the sfermion masses. To see this explicitly, note that the
gaugino masses are generated as a one loop effect, with the heavy
messenger(s) of SUSY breaking coupling directly to the gaugino. The
(generation blind) contributions to sfermion masses are two loop effects 
\cite{Gaume} through their coupling to gauginos. The only way for the
gauginos to communicate the SUSY breaking to the familon sector is through
the familon coupling to the sfermions making it a three loop effect with an
additional loop suppression which depends on the family symmetry gauge
coupling strength.

For low gauge mediation scale, $(m^{2}+\bar{m}^{2}/c)/\langle m_{\tilde{%
q}}^{2}\rangle $ needs to take values as small as $6 \times 10^{-4}.$ This is
possible if the family gauge symmetry coupling is very small, $\alpha
_{f}/4\pi <10^{-3}$. In practice one might expect a combination of the loop
factor and a mixing angle suppression below the maximum used in deriving the
upper bounds will allow for a solution with a larger gauge coupling.
Alternatively it may be that the gauge mediation scale is higher than $200$ TeV
leading to a further suppression of the bound compared to that shown in
Table \ref{ta:compare}.
For the case $c=-1$ the suppression is complete for the family symmetry gauge contribution because it is proportional to the square of the family charge (i.e. it generates $m^{2}=\bar{m}^{2}$).
In this case the bounds are satisfied for any value of the gauge coupling, as the non-gauge interactions are very heavily suppressed.

\section{Summary and conclusions \label{sec:Summary}}

To summarize, we have re-examined the bounds on continuous family symmetries
coming from the need to suppress the associated $D-$term contributions to
sfermion masses below the experimental bounds coming from FCNC processes.
The FCNC effects coming from the $D-$terms depend on unknown mixing angles
and we first derived upper bounds on these effects which are independent of
the mixing angles. We then compared these upper bounds with the experimental
bounds in each sector, accounting for the weakening of the constraints at
higher energy scales. For the case of gravity mediation the constraints are
only significant for the mixing of the first two generations. We identified
several ways in which these constraints are automatically satisfed without
appealing to a suppression involving alignment between the fermions and
sfermions. In the SUGRA and anomaly mediated cases, if the familon fields
spontaneously breaking the family symmetry relating the first two
generations have the same magnitude of family charge, the $D-$terms vanish
up to radiative corrections which may readily be within the constraints.
Even if the radiative corrections are large the $D-$terms may still be
within the limits if there is an underlying symmetry relating the couplings
of the familon fields and this may happen quite readily in family schemes
relying on the Froggatt-Nielsen mechanism to generate the fermion mass
hierarchy. Yet another possibility, motivated by orbifold string
compactified models, is that the familons have modular weights such that
they are anomalously light. This mechanism works irrespective of the family
charge carried by the familons. For the case of gauge mediated models the
lower mediation scale leads to stronger constraints which are non-trivial
for all the mixings between the three generations. For arbitrary familon charges these bounds can be satisfied for small family gauge coupling if, as is generally the case, the familons couple to the messenger sector only via the quark and lepton sector. For the case that the magnitude of familon charges are equal the bounds are satisfied for arbitrary gauge coupling.

In conclusion, the $D-$terms associated with continuous family symmetries
may be consistent with the experimental bounds on FCNC for a large class of
family models and SUSY breaking schemes. However in almost all cases the present
bounds on the mixing between the first two families are quite close to the
expected signals in these models demonstrating, yet again, the importance of
improving the experimental searches for FCNC effects.

\section*{Acknowledgements}

The work of I. de Medeiros Varzielas was supported by FCT under the grant
SFRH/BD/12218/2003. This work was partially supported by the EC 6th
Framework Programme MRTN-CT-2004-503369.


\begin{thebibliography}{99}
\bibitem{Masiero} %\cite{Gabbiani:1996hi}
F.~Gabbiani, E.~Gabrielli, A.~Masiero and L.~Silvestrini, 
%``A complete analysis of FCNC and CP constraints in general SUSY extensions
%of the standard model,''
Nucl.\ Phys.\ B \textbf{477}, 321 (1996) [arXiv:hep-ph/9604387]; 
%%CITATION = HEP-PH 9604387;%%
%\cite{Endo:2003te}
M.~Endo, M.~Kakizaki and M.~Yamaguchi, 
%``New constraint on squark flavor mixing from Hg-199 electric dipole
%moment,''
Phys.\ Lett.\ B \textbf{583}, 186 (2004) [arXiv:hep-ph/0311072]; 
%%CITATION = HEP-PH 0311072;%%
%\cite{Foster:2006ze}
J.~Foster, K.~i.~Okumura and L.~Roszkowski, 
% ``New constraints on SUSY flavour mixing in light of recent measurements at
%the Tevatron,''
arXiv:hep-ph/0604121. %%CITATION = HEP-PH 0604121;%%

\bibitem{Nir} Y.~Nir and N.~Seiberg, %``Should squarks be degenerate?,''
Phys.\ Lett.\ B \textbf{309} (1993) 337 [arXiv:hep-ph/9304307]; 
%%CITATION = HEP-PH 9304307;%%
%\cite{Cohen:1996vb}
A.~G.~Cohen, D.~B.~Kaplan and A.~E.~Nelson, 
%``The more minimal supersymmetric standard model,''
Phys.\ Lett.\ B \textbf{388} (1996) 588 [arXiv:hep-ph/9607394]. 
%%CITATION = HEP-PH 9607394;%%  

\bibitem{Kubo} %\cite{Kajiyama:2005rk}
Y.~Kajiyama, E.~Itou and J.~Kubo, 
%``Nonabelian discrete family symmetry to soften the SUSY flavor problem and to suppress proton decay,''
Nucl.\ Phys.\ B \textbf{743} (2006) 74 [arXiv:hep-ph/0511268]. 
%%CITATION = HEP-PH 0511268;%%

\bibitem{Murayama} %\cite{Kawamura:1994ys}
Y.~Kawamura, H.~Murayama and M.~Yamaguchi, 
% ``Low-Energy Effective Lagrangian In Unified Theories With Nonuniversal
%Supersymmetry Breaking Terms,''
Phys.\ Rev.\ D \textbf{51} (1995) 1337 [arXiv:hep-ph/9406245]; 
%%CITATION = HEP-PH 9406245;%%
%\cite{Murayama:1995fv}
H.~Murayama, 
%``Nonuniversal Susy Breaking, Hierarchy And Squark Degeneracy,''
arXiv:hep-ph/9503392. %%CITATION = HEP-PH 9503392;%%

\bibitem{Hall} %\cite{Hall:1985dx}
L.~J.~Hall, V.~A.~Kostelecky and S.~Raby, 
%``New Flavor Violations In Supergravity Models,''
Nucl.\ Phys.\ B \textbf{267} (1986) 415. %%CITATION = NUPHA,B267,415;%%

\bibitem{Ellis} %\cite{Ellis:1981tv}
  J.~R.~Ellis and D.~V.~Nanopoulos,
  %``Flavor Changing Neutral Interactions In Broken Supersymmetric Theories,''
  Phys.\ Lett.\ B {\bf 110}, 44 (1982);
  %%CITATION = PHLTA,B110,44;%%
  R.~Barbieri and R.~Gatto,
  %``Conservation Laws For Neutral Currents In Spontaneously Broken
  %Supersymmetric Theories,''
  Phys.\ Lett.\ B {\bf 110} (1982) 211;
  %%CITATION = PHLTA,B110,211;%%
%\cite{Inami:1982nu}
  T.~Inami and C.~S.~Lim,
  %``Natural Suppression Of Flavor Changing Neutral Currents In Supersymmetric
  %Gauge Theories,''
  Nucl.\ Phys.\ B {\bf 207}, 533 (1982).
  %%CITATION = NUPHA,B207,533;%%
J.~R.~Ellis, D.~V.~Nanopoulos and S.~Rudaz, %``Guts 3: Susy Guts 2,''
Nucl.\ Phys.\ B \textbf{202} (1982) 43. %%CITATION = NUPHA,B202,43;%%
%\cite{Hisano:1992jj}
J.~Hisano, H.~Murayama and T.~Yanagida, 
%``Nucleon decay in the minimal supersymmetric SU(5) grand unification,''
Nucl.\ Phys.\ B \textbf{402} (1993) 46 [arXiv:hep-ph/9207279]. 
%%CITATION = HEP-PH 9207279;%%
%\cite{Ramage:2003pf}
M.~R.~Ramage and G.~G.~Ross, %``Soft SUSY breaking and family symmetry,''
JHEP \textbf{0508} (2005) 031 [arXiv:hep-ph/0307389]. 
%%CITATION = HEP-PH 0307389;%%

\bibitem{Running} %\cite{Martin:1997ns}
S.~P.~Martin, %``A supersymmetry primer,''
arXiv:hep-ph/9709356; %%CITATION = HEP-PH 9709356;%%
%\cite{Allanach:2002nj}
  B.~C.~Allanach {\it et al.},
  %``The Snowmass points and slopes: Benchmarks for SUSY searches,''
in {\it Proc. of the APS/DPF/DPB Summer Study on the Future of Particle Physics (Snowmass 2001) } ed. N.~Graf,
{\it In the Proceedings of APS / DPF / DPB Summer Study on the Future of Particle Physics (Snowmass 2001), Snowmass, Colorado, 30 Jun - 21 Jul
2001, pp P125}
  [arXiv:hep-ph/0202233];
  %%CITATION = HEP-PH 0202233;%%
  W.~Porod,
  %``SPheno, a program for calculating supersymmetric spectra, SUSY particle
  %decays and SUSY particle production at e+ e- colliders,''
  Comput.\ Phys.\ Commun.\  {\bf 153} (2003) 275
  [arXiv:hep-ph/0301101].
  %%CITATION = HEP-PH 0301101;%%

\bibitem{ivo}
S.~F.~King and G.~G.~Ross, 
%``Fermion masses and mixing angles from SU(3) family symmetry and
%unification,''
Phys.\ Lett.\ B \textbf{574} (2003) 239 [arXiv:hep-ph/0307190];
I.~de Medeiros Varzielas and G.~G.~Ross, 
%``SU(3) family symmetry and neutrino bi-tri-maximal mixing,''
Nucl.\ Phys.\ B \textbf{733} (2006) 31 [arXiv:hep-ph/0507176];
%\cite{deMedeirosVarzielas:2006fc}
  I.~de Medeiros Varzielas, S.~F.~King and G.~G.~Ross,
  %``Neutrino tri-bi-maximal mixing from a non-Abelian discrete family
  %symmetry,''
  arXiv:hep-ph/0607045.
  %%CITATION = HEP-PH 0607045;%%

\bibitem{casas} 
J.~A.~Casas, G.~B.~Gelmini and A.~Riotto, 
%``F-term inflation in superstring theories,''
Phys.\ Lett.\ B \textbf{459} (1999) 91 [arXiv:hep-ph/9903492].

\bibitem{Gaume}
L.~Alvarez-Gaume, M.~Claudson and M.~B.~Wise, 
%``Low-Energy Supersymmetry,''
Nucl.\ Phys.\ B \textbf{207}, 96 (1982); %%CITATION = NUPHA,B207,96;%%
%\cite{Giudice:1997ni}
G.~F.~Giudice and R.~Rattazzi, 
%``Extracting Supersymmetry-Breaking Effects From Wave-Function
%Renormalization,''
Nucl.\ Phys.\ B \textbf{511}, 25 (1998) [arXiv:hep-ph/9706540]; 
%%CITATION = HEP-PH 9706540;%%
%\cite{Giudice:1998bp}
G.~F.~Giudice and R.~Rattazzi, 
%``Theories with gauge-mediated supersymmetry breaking,''
Phys.\ Rept.\ \textbf{322}, 419 (1999) [arXiv:hep-ph/9801271]. 
%%CITATION = HEP-PH 9801271;%%

\end{thebibliography}
\end{document}